\newcommand{\diracslash}[1]{#1\llap{/\kern2pt}}

\newcommand{\be}{\begin{equation}}
\newcommand{\ee}{\end{equation}}
\newcommand{\bea}{\begin{eqnarray}}
\newcommand{\eea}{\end{eqnarray}}
\newcommand{\ba}[1]{\begin{array}{#1}}
\newcommand{\ea}{\end{array}}

\documentclass{ws-procs9x6}
\usepackage{graphicx}
\usepackage{amssymb}
\usepackage{amsmath}
\usepackage{multirow}
\begin{document}

\title
{Kinetics of Chiral Phase Transitions in Quark Matter} 

\author {Awaneesh Singh, Sanjay Puri} 
\address{School of Physical Sciences, Jawaharlal Nehru University,
 New Delhi -- 110067, India. }
\author {Hiranmaya Mishra } 
\address{Theory Division, Physical Research Laboratory, Navrangpura,
 Ahmedabad -- 380009, India.\\
E-mail: hm@prl.res.in}

\def\be{\begin{equation}}
\def\ee{\end{equation}}
\def\bearr{\begin{eqnarray}}
\def\eearr{\end{eqnarray}}
\def\zbf#1{{\bf {#1}}}
\def\bfm#1{\mbox{\boldmath $#1$}}
\def\hf{\frac{1}{2}}

\begin{abstract}
We study the kinetics of chiral transitions in quark matter using a microscopic framework (Nambu-Jona-Lasinio model) and a phenomenological model (Ginzburg-Landau free energy). We focus on the coarsening dynamics subsequent to a quench from the massless quark phase to the massive quark phase. The morphology of the ordering system is characterized by the scaling of the order-parameter correlation function. The domain growth process obeys the Allen-Cahn growth law, $L(t)\sim t^{1/2}$. We also study the growth of bubbles of the stable massive phase from the metastable massless phase.
\end{abstract}
\bodymatter
The nature of the QCD phase diagram in the plane of temperature ($T$) and baryon chemical potential ($\mu$) has been studied extensively over the last few decades \cite{rischkerev,rajgopalrev}. For $\mu=0$, finite-temperature perturbative QCD calculations have been complemented by first-principle calculations like lattice QCD simulations \cite{karschlect,laerphill,laersau}. However, for $\mu \ne 0$, lattice QCD calculations are limited to small values of $\mu$ \cite{latmu}. In QCD with two massless quarks, the equilibrium chiral phase transition is expected to be a second-order transition at zero baryon densities. However, in nature, the light quarks are not exactly massless and the second-order phase transition is replaced by an analytical crossover. This picture is consistent with lattice QCD simulations with a transition temperature $T_c\sim 140 - 190$ MeV \cite{Tclattice}. Calculations based on different effective potentials, on the other hand, indicate the phase transition to be first-order at large $\mu$ and small T. This means that the phase diagram will have a tri-critical point, where the first-order chiral transition becomes second-order (for vanishing quark masses) or ends (for non-vanishing quark masses) at a critical end point in the phase diagram. The location of the tri-critical point (TCP)($\mu^E$, $T^E$) in the phase digram has been estimated by lattice QCD simulations as $\mu^E/T_c\simeq0.94$ and $\mu^E/T^E\simeq 1.8$ \cite{gavaigupta}.

Heavy-ion collision experiments at high energies produce hot and dense strongly-interacting matter, and provide the opportunity to explore the phase diagram of QCD. While the high-$T$ and small-$\mu$ region of the QCD phase diagram has been explored in recent experiments. Future heavy-ion collision experiments plan to explore the high baryon density regime, particularly the region around the tricritical point \cite{cpod}. The experiments at Relativistic Heavy Ion Collider (RHIC) provide clear signals that the nuclear matter undergoes a phase transition to partonic phases at sufficiently large value of the energy density. However, the nature of the phase transition still remains an open question. We might mention here that lattice QCD thermodynamics has an built in equilibrium assumption while the heavy ion experiments are essentially nonequilibrium processes. It is worthwhile to mention here that, in a phase transition process, information about which equilibrium phase has lowest free energy is not sufficient to discuss all possible structures that the system can have. One has to understand the kinetics of the process by which the phase ordering or disordering proceeds and the nature of nonequilibrium structures that the system must go through on its way to reach equilibrium.

It is not surprising, therefore that much attention has been focused on dynamics of chiral transition in dense quark matter, particularly near the critical point. The critical behavior and the fluctuations of conserved charges in the presence of spinodal decomposition in the context of chiral transition have been explored in Nambu-Jona-Lasinio (NJL) model \cite{sasaki}.  There have also been preliminary studies \cite{dumitru} of the kinetics of the chiral phase transition in a nonlinear sigma model coupled to quarks. In this context, the effect of dissipation in slowing down spinodal decomposition has also been studied in this model \cite{fragaplb}. The dynamics of first-order phase transitions has been considered recently in Ref. \cite{fraga}. Simulated by the findings of lattice QCD, few studies have also been considered for the case of smooth crossover, its dynamics as compared to a strongly first order transition\cite{stoeckeretc}. Further, a Langevin equation has been derived within NJL model describing the behavior of the fluctuation above the critical temperature at finite chemical potential within a linear approximation\cite{tomoinpa}.

On the otherhand, there has been intense research interest in the kinetics of phase transitions, and the phase ordering process that occurs after a rapid quench in system parameters e.g., temperature, pressure \cite{pw09, aj94} in various condensed matter systems. During the transition, the system develops a spatial structure of randomly-distributed domains which coarsen with time. This domain growth process has been extensively studied in many condensed matter systems like ferromagnets, binary fluids, liquid crystals, etc.

In the present work, we focus on the kinetics of the chiral phase transition subsequent to a quench from the disordered phase (with vanishing quark condensate) to the ordered phase for quark matter. This means while the equilibrium configuration at zero temperature is an ordered state, the system finds itself in a configuration from the ensemble appropriate to a high temperature. We are thus interested in the far-from-equilibrium evolution of the system and pattern dynamics associated with it.

To model chiral symmetry breaking in QCD, we use the two-flavor Nambu-Jona-Lasinio (NJL) model \cite{klevansky, glfree} with the Hamiltonian
\begin{eqnarray}
{\cal H} = \sum_{i,a}\psi^{ia \dagger}\left(-i\vec{\alpha}\cdot\vec{\nabla} + \gamma^0 m_i \right)\psi^{ia}-G\left[(\bar\psi\psi)^2-(\bar\psi\gamma^5 \vec{\tau} \psi)^2\right].
\label{ham}
\end{eqnarray}
Here, $m_i$ is the current quark mass -- we take this to be the same ($m_i=m$) for both u and d quarks. The parameter $G$ denotes the quark-quark interaction strength. Further, $\tau$ is the Pauli matrix acting in flavor space. The quark operator $\psi$ has two indices $i$ and $a$, denoting theflavor and color indices, respectively. This model exhibits a second-order chiral  phase transition for massless quarks at small $\mu$ and high $T$, and a first-order transition at large $\mu$ and small $T$.

To describe the ground state, we take an ansatz with quark-antiquark condensates \cite{hmnj}:
\begin{equation} 
|vac\rangle= \exp\!\!\left[\!\int\!\! d\vec{k}~ q_I^{0i}(\vec{k})^\dagger(\vec{\sigma}\cdot\vec{k})h_i(\vec{k})\tilde q_I^{0i} (-\vec{k})-\mathrm{h.c.}\right]\!|0\rangle.
\label{uq}
\end{equation}
Here, $q^\dagger$, $\tilde q$ are two-component quark and antiquark creation operators, and $|0\rangle$ is the perturbative chiral vacuum. Further, $h_i(\vec{k})$ is a variational function related to the quark-antiquark condensate as
\begin{align}
\langle\bar{\psi}\psi\rangle = -\frac{3}{\pi^3} \sum_{i=1}^2 \int \!d\vec{k}\,\sin[2h_i(\vec{k})].
\label{qc}
\end{align}
This flavor-dependent function can be determined by minimizing the energy at $T=0$, or the thermodynamic potential at nonzero $T$ and density. In the mean-field approximation and, near the chiral phase transition, we write down the expression for the thermodynamic potential as 
\begin{align}
\tilde\Omega(M,\beta,\mu) =& \;-\dfrac{12}{(2\pi)^3\beta}\displaystyle\int \! d\vec{k}\;\Big\{ \ln\left[1+e^{-\beta\left( \sqrt{k^2+M^2} - \mu \right)}\right] \notag\\
& \qquad\qquad\qquad\quad + \ln\left[1+e^{-\beta\left( \sqrt{k^2+M^2} + \mu \right)}\right] \Big\}   \notag\\
&\; -\dfrac{12}{(2\pi)^3}\displaystyle\int \! d\vec{k} \; \left(\sqrt{\vec{k}^2+M^2}-k\right)+ \dfrac{M^2}{4G}.
\label{tomega}
\end{align}
Here, we have taken vanishing current quark mass, and introduce $M=-2g\rho_s$ with $\rho_s=\langle\bar\psi\psi\rangle$ being the scalar density and $g=G[1+1/(4N_c)]$. The details of the mean-field approximation are reported elsewhere \cite{awaneesh1}.

\begin{figure}[!hb]
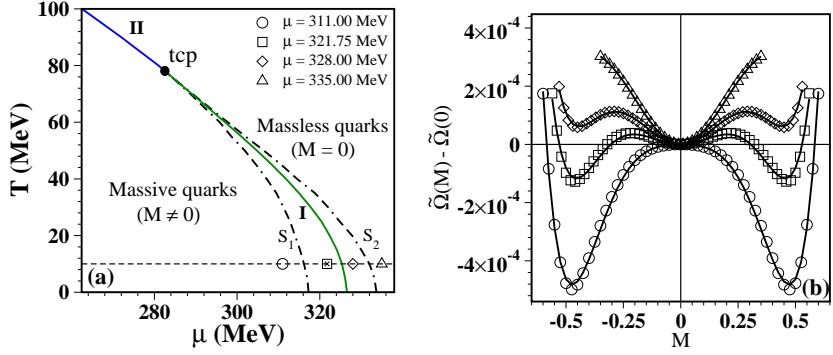

\centering
\begin{tabular}{c c }
\includegraphics[width=0.45\textwidth]{fig1a.eps}&
\includegraphics[width=0.47\textwidth]{fig1b.eps}\\
\end{tabular}
\caption{(a) Phase diagram of the Nambu-Jona-Lasinio (NJL) model in the ($\mu, T$)-plane for zero current quark mass. A line of first-order transitions (I, green online) meets a line of second-order transitions (II, blue online) at the tricritical point (tcp). We have $(\mu_\text{tcp}, T_\text{tcp}) \simeq (282.58, 78)$ MeV. The dot-dashed lines $S_1$ and $S_2$ denote the spinodals or metastability limits for the first-order transitions. The open symbols denote 4 combinations of $\left(\mu, T\right)$, chosen to represent qualitatively different shapes of the NJL potential. The cross denotes the point at which we quench the system for $b<0$. (b) Plot of $\tilde\Omega\left(M, \beta, \mu \right)$ from Eq. (\ref{tomega}) as a function of $M$. The ($\mu, T$)-values are marked in (a).
The solid lines superposed on the potentials correspond to the GL potential in Eq. (\ref{p6}) with $a$ from Eq. (\ref{coff}), and $b$, $d$ being fit parameters (cf. Table \ref{tab}).}
\label{fig1}
\end{figure}

The phase diagram for the chiral transition in the ($\mu,T$)-plane resulting from Eq. (\ref{tomega}) is shown in Fig. \ref{fig1}(a). For the numerical calculation of the thermodynamic potential, we have taken here a three-momentum ultraviolet cutoff $\Lambda=653.30$ MeV, and the four-fermion coupling $G=5.0163\times 10^{-6}$ $\mathrm{MeV^{-2}}$ \cite{askawa}. With these parameters, the vacuum mass of quarks is $M\simeq 312$ MeV. At $T=0$, a first-order transition takes place at $\mu \simeq 326.321$ MeV. For $\mu=0$, a second-order transition takes place at $T \simeq 190$ MeV. The first-order line (I) meets the second-order line (II) at the tricritical point $(\mu_\text{tcp},T_\text{tcp})\simeq(282.58, 78)$ MeV. The first-order transition is characterized by the existence of metastable phases. The limit of metastability is denoted by the dashed lines $S_1$, $S_2$ in Fig. \ref{fig1}(a), referred to as spinodal lines.

Close to the phase boundary, the potential in Eq. (\ref{tomega}) may be expanded as a Ginzburg-Landau (GL) potential in the order parameter $M$: 
\begin{equation}
\tilde\Omega\left(M \right)= \tilde\Omega\left(0 \right) + \frac{a}{2}M^2 + \frac{b}{4}M^4 + \frac{d}{6}M^6 + O(M^8),
\label{p6}
\end{equation}
correct upto logarithmic corrections \cite{sasaki,glfree}. In the following, we consider the expansion of potential $\tilde\Omega\left(M \right)$ upto the $M^6$-term. This will prove adequate to recover the phase diagram in Fig. \ref{fig1}(a), as we see shortly. The first two coefficients in Eq. (\ref{p6}) can be obtained by comparison with Eq. (\ref{tomega}) as
\begin{align}
\tilde\Omega(0) =&\;-\dfrac{6}{\pi^2\beta}\displaystyle\int_0^\Lambda \!\!\! dk\,\, k^2 \left\lbrace  
\ln\left[1+e^{-\beta(k-\mu)}\right] + \ln\left[1+e^{-\beta(k+\mu)}\right]\right\rbrace, \nonumber \\
a =& \; \dfrac{1}{2G} - \dfrac{3\Lambda^2}{\pi^2} + \dfrac{6}{\pi^2}\displaystyle\int_0^\Lambda \!\!\! dk\,\,k\left[ \dfrac{1}{1+e^{\beta(k-\mu)}} + \dfrac{1}{1+e^{\beta(k+\mu)}}\right]. 
\label{coff}
\end{align}
We treat the higher coefficients as phenomenological parameters, which are obtained by fitting $\tilde\Omega\left(M \right)$ in Eq. (\ref{p6}) to the integral expression for $\tilde\Omega$ in Eq. (\ref{tomega}). There are two free parameters in the microscopic theory ($\mu$ and $T$), so we consider the $M^6$-GL potential with parameters $b$ and $d$. For stability, we require $d>0$.

In Fig. \ref{fig1}(b), we plot $\tilde\Omega\left(M \right)-\tilde\Omega\left(0\right)$ from Eq. (\ref{tomega}) as a function of $M$. We show 4 combinations of $\left(\mu, T\right)$ as marked in Fig. \ref{fig1}(a), chosen to represent qualitatively different shapes of the potential. The solid lines superposed on the data sets in Fig. \ref{fig1}(b) correspond to the GL potential in Eq. (\ref{p6}) with $a$ from Eq. (\ref{coff}), and $b$, $d$ being fit parameters (see Table \ref{tab}). 
\begin{table}[!ht]
\begin{center}
\renewcommand{\tabcolsep}{0.35cm}
\renewcommand{\arraystretch}{1.5}
    \begin{tabular}{| c || c | c | c | c |}
    \hline
    \multicolumn{5}{|c|}{T=10 MeV} \\ \hline
    $\mu$ (MeV) & $a/\Lambda^2$ & $b$ & $d\Lambda^{2}$ & $\lambda=|a|d/b^2$ \\ \hline
    311.00 & -1.305$\times 10^{-3}$ &  0.0924 & 0.439 & 0.067 \\ 
    321.75 &  3.539$\times 10^{-3}$ & -0.101 & 0.402 & 0.140 \\ 
    328.00 &  6.431$\times 10^{-3}$ & -0.111& 0.396 & 0.206 \\
    335.00 &  9.736$\times 10^{-3}$ & -0.101& 0.265 & 0.255 \\ \hline
    \end{tabular}
\end{center}
\caption{The coefficients ($a$, $b$, $d$) of the GL-potential for 4 different values of $\mu$ at $T=10$ MeV. The coefficients $a$, $b$ and $d$ are measured in units of $\Lambda^2$, $\Lambda^0$ and $\Lambda^{-2}$ respectively, where $\Lambda=653.30$ MeV.}
\label{tab}
\end{table}

\begin{figure}[!htb]
\centering
\includegraphics[width=0.7\textwidth]{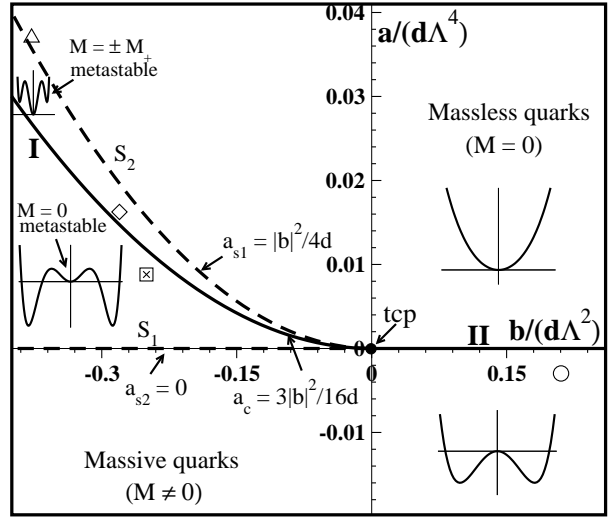}
\caption{Phase diagram in ($b/(d\Lambda^2)$, $a/(d\Lambda^4)$)-space for the GL-free energy in Eq. (\ref{p6}). The typical forms of the GL potential in various regions are shown in the figure. The open symbols denote the ($\mu, T$)-values marked in Fig. \ref{fig1}(a). The cross denotes the point where we quench the system for $b<0$, shows all different possible quenches.}
\label{fig2}
\end{figure}

The extrema of the potential in Eq. (\ref{p6}) are determined by the gap equation $f'(M)=aM+bM^3+dM^5=0$. The solutions are $ M=0$, and $M_{\pm}^2= (-b\pm \sqrt{b^2 -4ad})/(2d)$. For $b>0$, the transition is second-order, analogous to an $M^4$-potential -- the stationary points are $M=0$ (for $a>0$) or $M=0$, $\pm M_+$ (for $a<0$). For $a<0$, the preferred equilibrium state is the one with massive quarks. For $b<0$, the solutions of the gap equation are as follows: (i) $M=0$ for $a>b^2/(4d)$, (ii) $M=0$, $\pm M_+$, $\pm M_-$ for $b^2/(4d)>a>0$, and (iii) $M=0$, $\pm M_+$ for $a<0$. A first-order transition takes place at $a_c=3b^2/(16d)$ with the order parameter jumping discontinuously from $M=0$  to $M=\pm M_+=\pm (3|b|/4d)^{1/2}$. The phase diagram for the GL potential is shown in Fig. \ref{fig2}. The tricritical point is located at $b_\text{tcp}=0$, $a_\text{tcp}=0$ [cf. Fig. \ref{fig1}(a)]. The 4 combinations of $\left(\mu, T\right)$-values marked in Fig. \ref{fig1}(a) are identified with the same symbols in Fig. \ref{fig2}.

Next, we study dynamical problems in the context of the above free energy. Consider the dynamical environment of a heavy-ion collision. As long as the evolution is slow compared to the typical re-equilibration time, the order parameter field will be in local equilibrium. We consider a system which is rendered thermodynamically unstable by a rapid quench from the massless phase to the massive phase in Figs. \ref{fig1} or \ref{fig2}. The unstable disordered state evolves via the emergence and growth of domains rich in the preferred phase \cite{pw09, aj94}. The coarsening system is inhomogeneous, and we account for this by including a surface tension term in the GL free energy:
\begin{equation}
\Omega[M]=\int d\vec{r} \left[\frac{a}{2} M^2+\frac{b}{4} M^4+\frac{d}{6} M^6
+\frac{K}{2}\left(\vec{\nabla} M\right)^2\right].
\label{omgl}
\end{equation}
In Eq. (\ref{omgl}), $\Omega[M]$ is a functional of the spatially-dependent order parameter $M(\vec{r})$, and $K$ measures the energy cost of spatial inhomogeneities, i.e., surface tension.

The evolution of the system is described by the time-dependent Ginzburg-Landau (TDGL) equation: 
\begin{equation}
\frac{\partial}{\partial t}M\left(\vec{r},t\right)= -\Gamma 
\frac{\delta \Omega\left[M \right]}{ \delta M}+\theta\left(\vec{r},t\right), 
\label{ke}
\end{equation}
which models the over-damped relaxational dynamics of $M(\vec{r},t)$ to the minimum of $\Omega\left[M \right]$ \cite{hohenrev}. Here, $\Gamma$ is the inverse damping coefficient, and $\theta(\vec{r},t)$ is the noise term satisfying the fluctuation-dissipation relation: $\left\langle\theta\left(\vec{r},t\right) \right\rangle = 0 $ and  $\left\langle \theta(\vec{r'},t')\theta(\vec{r''},t'') \right\rangle = 2\Gamma T\delta(\vec{r'}-\vec{r''})\delta\left(t'-t''\right)$. We use the natural scales of order parameter, space and time to introduce dimensionless variables: $M=M_0M'$ ($M_0=\sqrt{|a|/|b|}$); $\vec{r}=\xi\vec{r'}$ ($\xi=\sqrt{K/|a|}$); $t=\tau t'$ $(\tau=(\Gamma|a|)^{-1})$; $\theta= (\Gamma|a|^{3/2}T^{1/2}/|b|^{1/2})~\theta'$. Dropping the primes, we obtain the dimensionless TDGL equation: 
\begin{eqnarray}
\frac{\partial}{\partial t}M\left(\vec{r},t\right)=-\mathrm{sgn} \left(a\right)M - \mathrm{sgn}\left(b\right)M^3-\lambda M^5 + \nabla^2 M +\theta\left(\vec{r},t\right), 
\label{ke2} 
\end{eqnarray}
where $\lambda=|a|d/b^2 >0$. For $T=10$ MeV, as $\mu$ takes values (in MeV) 311, 321.75, 328 and 335, the corresponding values of $\lambda$ are 0.067, 0.14, 0.206 and 0.255 from Table \ref{tab}.

The following evolutions are performed in the dimensionless units of length and time. To get them in physical units, one has to multiply them with the appropriate dimensional parameters. Two unknown parameters are $K$, the strength of the surface tension and the inverse damping coefficient $\Gamma$. These parameters are obtained from the GL coefficients (see Table \ref{tab}). Let us note that the surface energy can be calculated to be $\sigma = \sqrt{K}(a^{3/2}/b)\int dz(dM/dz)^2$. The surface energy ($\sigma$) for quark matter is poorly known and varies from 10-100 MeV/$\text{fm}^2$ at small temperatures \cite{hc93}. To get an idea about the length scale we can take $\sigma=50$ MeV/$\text{fm}^2$. For $T=10$ MeV, we can estimate $\sqrt{K/a} = 1.6$ fm. This factor has to be multiplied with the dimensionless length scale to get the physical spatial length scale in units of fermi. Similarly for the inverse damping coefficient $\Gamma$, we take it to be of the order $2T/s$, where $s$ is a quantity of order 1 \cite{kk92}. This leads to $t=2.6$ $t'$ fm/$s$.

In the following, we consider the phase transition kinetics for two different quench possibilities. The first case corresponds to high $T$ and low baryon density ($\mu$), where the quenching is done through the second-order line (II) in Fig. \ref{fig1}(a) or Fig. \ref{fig2}. The second case corresponds to low $T$ and high baryon density ($\mu$), where the phase conversion process can probe the metastable region of the phase diagram. This can be  achieved by quenching through the first-order line (I) in Fig. \ref{fig1}(a) or Fig. \ref{fig2}. Note that both these scenarios can be studied using Eq. ({\ref{ke2}}) by choosing a fixed value of $\lambda$ and appropriate values of $a$, $b$. In our simulation, we have used $\lambda=0.14$. This value of $\lambda$ corresponds to, e.g., ($\mu$, $T$)= ($231.6$ MeV, $85$ MeV) or ($321.75$ MeV, $10$ MeV) in Fig. \ref{fig1}(a).

First, we focus our attention to the ordering dynamics in the context of the phase diagram of Fig. \ref{fig2} for the case of $b>0$. This corresponds to the second order transition case relevant for low chemical potential regime in the context of the phase diagram of Fig. \ref{fig1}(a). For $b>0$, the chiral transition occurs when $a<0$. We solve Eq. (\ref{ke2}) numerically using an Euler-discretization scheme with an isotropic Laplacian. We have implemented it on a $d$=3 lattice of  size $N^3$ ($N=256$), with periodic boundary conditions in all directions. The dimensionless mesh sizes are $\Delta x=1.0$ and $\Delta t=0.1$, which satisfy the numerical stability condition. We have further confirmed that the spatial mesh size is sufficiently small to resolve the interface region \cite{op86}. 

\begin{figure}[!htb]
\centering
\includegraphics[width=0.9\textwidth]{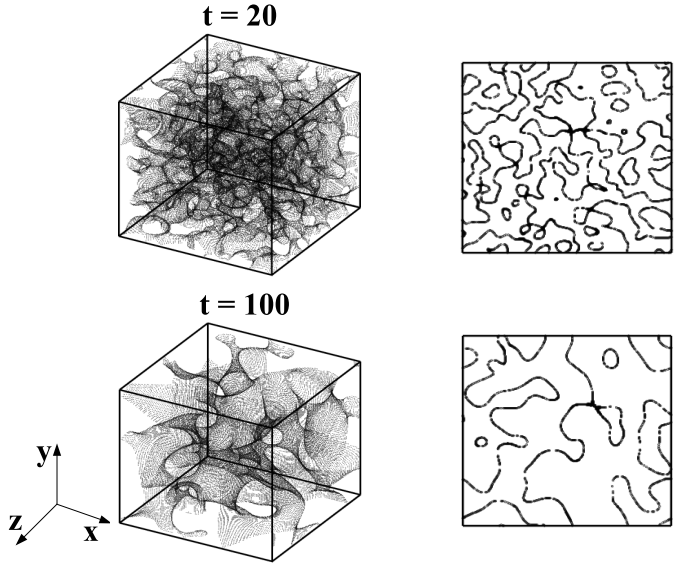}
\caption{Interface evolution after temperature quench through 
second-order line (II) in Figs. \ref{fig1}(a) or \ref{fig2}.
 The $d$=3 snapshots on the left show the interfaces 
($M=0$) at $t=20$, $100$ (in units of $\tau$). They were obtained by 
numerically solving Eq. (\ref{ke2}) as described in the text with $a<0$, $b>0$, $\lambda=0.14$. The noise amplitude was $\epsilon=0.008$. The frames on the right show a cross-section of the snapshots at $z=N/2$.}
\label{fig3}
\end{figure}

Figure \ref{fig3} shows the evolution of a disordered initial condition for Eq. (\ref{ke2}) with $b>0$ and $a<0$. It corresponds to a temperature quench through the second-order line (II) in Figs. \ref{fig1}(a) or \ref{fig2}. The initial state consists of small-amplitude thermal fluctuations about the massless phase $M=0$. The system rapidly evolves into domains of the massive phase with $M\simeq M_+$ and $M\simeq-M_+$. The interfaces between these massive domains correspond to $M=0$. The interface evolution is shown in the snapshots (frames on left) of Fig. \ref{fig3}. The frames on the right show the interface structure in a cross-section of the snapshots.

\begin{figure}[!htb]
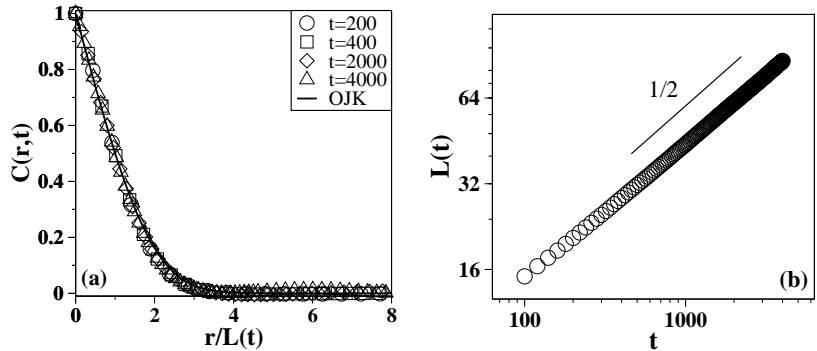

\centering
\begin{tabular}{c c }
\includegraphics[width=0.45\textwidth]{fig4a.eps}&
\includegraphics[width=0.45\textwidth]{fig4b.eps}\\
\end{tabular}
\caption{(a) Scaling of order-parameter correlation function, $C(r,t)$ vs. $r/L$, for $\lambda=0.14$ at four different times. The different data sets collaps onto a master curve. The statistical data is obtained as an average over 10 independent runs on $4096^2$ lattices. The length scale $L(t)$ is defined as the distance over which $C(r,t)$ decays from $1 \rightarrow 1/2$. The solid line denotes the OJK result \cite{ojk82}. (b) Time-dependence of domain size, $L(t)$ vs. $t$, for $\lambda=0.14$. The coarsening process obeys the Allen-Cahn (AC) growth law, $L(t)\sim t^{1/2}$.}
\label{fig4}
\end{figure}

The characteristic length scale $L(t)$ of the domains grows with time. The growth process is analogous to coarsening in the TDGL equation with an $M^4$-potential, where the coarsening is driven by kinks with the equilibrium profile $M(z)=\tanh(\pm z/\sqrt 2)$ \cite{pw09, aj94}. The order-parameter correlation function $C(r,t)$ shows \emph{dynamical scaling} $C(r,t)=f(r/L)$. The scaling function $f(x)=(2/\pi)\sin^{-1}(e^{-x^2})$ has been calculated by Ohta \emph{et al.} (OJK) \cite{ojk82} in the context of an ordering ferromagnet. In Fig. \ref{fig4}(a), we demonstrate that $C(r,t)$ shows dynamical scaling for the evolution in Fig. \ref{fig3} shows the scaling property. Further, the domain scale obeys the Allen-Cahn (AC) growth law, $L(t)\sim t^{1/2}$ \cite{aj94} [see Fig. \ref{fig4}(b)]. Typically, the interface velocity $v\sim dL/dt \sim 1/L$, where $L^{-1}$ measures the local curvature of the interface. This yields the AC growth law. The same growth law has also been obtained via a closed time path formalism of relativistic finite-temperature field theory applied to the NJL model \cite{das}.

Next, let us consider the case with $b<0$ in Fig. \ref{fig2}. In this case, a first-order chiral transition occurs for $a<a_c=3b^2/(16d)$ (or $\lambda<\lambda_c=3/16)$. For $a<0$, the potential has a double-well structure and the ordering dynamics is equivalent to $M^4$-theory, i.e., the domain growth scenario is similar to Figs. \ref{fig3} and \ref{fig4}. We focus on a quench from the disordered state (with $M=0$) to $0<a<a_c$ or $0<\lambda<\lambda_c$, corresponding to a quench between the first-order line (I) and $\mathrm{S_1}$ in Figs. \ref{fig1}(a) or \ref{fig2} -- the corresponding points are denoted by crosses. The massless state ($M=0$) is now a metastable state of the $M^6$-potential. The chiral transition proceeds via the nucleation and growth of droplets of the preferred phase ($M=\pm M_+$). The nucleation results from large fluctuations in the initial condition or thermal fluctuations during the evolution. 
\begin{figure}[!htb]
\centering
\includegraphics[width=0.85\textwidth]{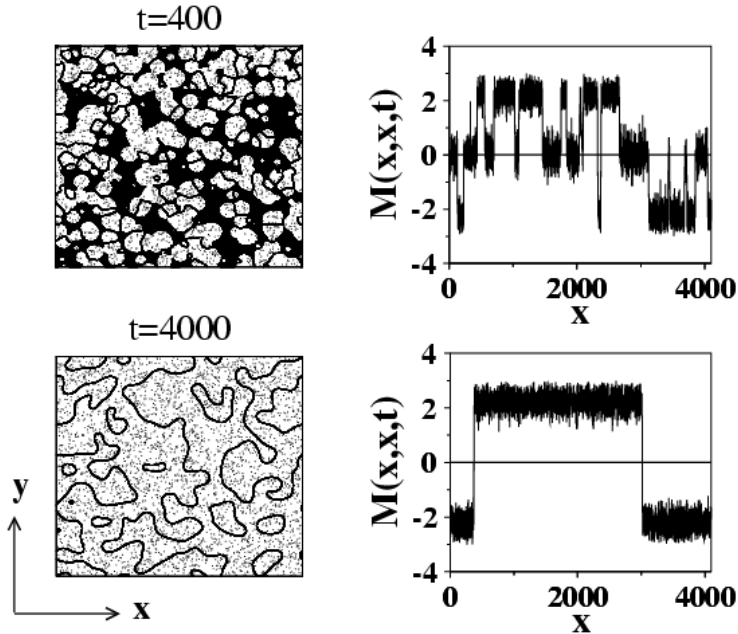}
\caption{Interface evolution after shallow temperature quench through first-order line (I) in Figs. \ref{fig1}(a) or \ref{fig2}. The $d$=2 snapshots on the left show the interfaces ($M=0$) at $t=400$, $4000$. They were obtained by solving Eq. (\ref{ke2}) with $b<0$, $a_c>a>0$ and $\lambda=0.14$. The frames on the right show the variation of the order parameter along the diagonal. Notice that the metastable patches ($M=0$) at $t=400$ are absent at later times.}
\label{fig5}
\end{figure}

In Fig. \ref{fig5}, we show the nucleation and growth process. At early times ($t=400$), the system is primarily in the $M=0$ phase with small droplets of the preferred phase. These droplets grow in time and coalesce into domains. The subsequent coarsening of these domains is analogous to that in Figs. \ref{fig3}-\ref{fig4}, through the interfacial kinks are slightly flatter in the $M\simeq0$ region. In the late stages of growth, there is no memory of the nucleation dynamics which characterized growth during the early stages.

\begin{figure}[!htb]
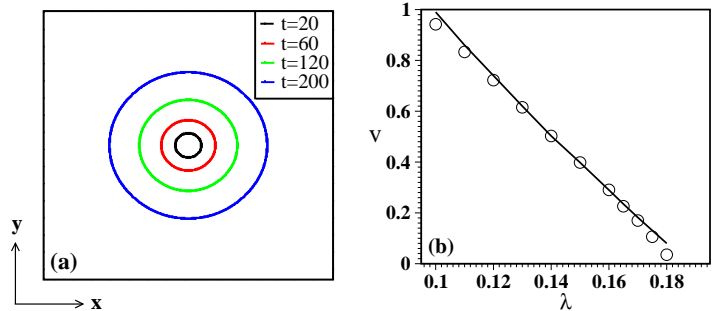

\centering
\begin{tabular}{c c }
\includegraphics[width=0.38\textwidth]{fig6a.eps}&
\includegraphics[width=0.40\textwidth]{fig6b.eps}\\
\end{tabular}
\caption{(a) Growth of droplet of the preferred phase ($M = +M_+$) in a background of the metastable phase ($M=0$) for $b<0$, $a_c>a>0$ and $\lambda = 0.14$. We show the boundary of the droplet at four different times. The inner circle corresponds to a droplet at time $t=20$. (b) Droplet velocity $v$ vs. $\lambda$. The circles refer to numerical data, while the solid line corresponds to the result from a phase-plane analysis \cite{awaneesh1}.}
\label{fig6}
\end{figure}

For $b<0$ and $0<a<a_c$, we have also studied the growth of single droplets of the preferred phase ($M=+M_+$) in a background of the metastable phase ($M=0$) for $\lambda=0.14$ [see Fig. \ref{fig6}(a)]. We start with an initial configuration of a bubble of radius $R_0>R_c$ such that $M(r)=M_+$ for $r<R_0$ and $M(r)=0$ for $r>R_0$, where $R_c$ is the critical size of the droplet. This configuration is evolved as per Eq. (\ref{ke2}) with the noise term. The noise term is taken sufficiently large to overcome the metastable barrier. The droplets have a unique growth velocity $v(\lambda)$, which depends on the degree of undercooling ($\lambda$): $v \rightarrow 0$ as $\lambda \rightarrow \lambda_c^-$. We have obtained $v(\lambda)$ by undertaking a phase-plane analysis of the traveling-wave solutions of Eq. (\ref{ke2}) with $\theta =0$. The droplet interface corresponds to a saddle connection between the fixed points $+M_+ \rightarrow 0$ or $-M_+ \rightarrow 0$. The details of this analysis will be presented later \cite{awaneesh1}. In Fig. \ref{fig6}(b), we plot numerical results for $v(\lambda)$ vs. $\lambda$ along with our theoretical result.

Before concluding, we briefly discuss the inertial counterpart of the overdamped TDGL equation (\ref{ke}). The TDGL equation contains first-order time-derivatives and is not Lorentz invariant. This is reasonable as we are considering finite-temperature field theory, where there is no Lorentz invariance. However, in some situations, the inertial terms can also play an important role in the evolution dynamics. The inertial TDGL equation has the following dimensionless form:
\begin{eqnarray}
\frac{\partial^2M}{\partial t^2} + \gamma\frac{\partial M}{\partial t}=
-\mathrm{sgn}\left(a\right)M - \mathrm{sgn}\left(b\right)M^3- \lambda M^5 +\nabla^2M +\theta\left(\vec{r},t\right),
\label{ke1}
\end{eqnarray}
where $\gamma$ measures the relative strengths of the damping and inertial terms. In principle, Eq. (\ref{ke1}) can be obtained from a microscopic field-theoretic description of the nonequilibrium dynamics of the scalar field at finite temperatures \cite{ramos}. We have also studied ordering dynamics for the inertial case, and will present details of our results later \cite{awaneesh1}. Here, we mention the main results  of our study. The ordering dynamics in the inertial case is analogous to that in the overdamped case, except that nucleation does not have a significant effect even during the early stages of evolution for quenches to $b<0$ and $0<a<a_c$. The droplets grow very rapidly and merge to form a bicontinuous domain structure characteristic of late-stage domain growth. The domain growth law is again the AC law, $L(t)\sim t^{1/2}$.

To summarize: we have studied the kinetics of chiral phase transitions in QCD. In terms of the quark degrees of freedom, the phase diagram is obtained using the NJL model. An equivalent coarse-grained description is obtained from a $M^6$-Ginzburg-Landau (GL) free energy. The chiral kinetics is modeled via the TDGL equation, and we consider both the overdamped and inertial cases. We study the ordering dynamics resulting from a sudden temperature quench through the first-order (I) or second-order (II) transition lines in Figs. \ref{fig1}(a) or \ref{fig2}. For quenches through II and deep quenches through I, the massless phase is spontaneously unstable and evolves to the massive phase via spinodal decomposition. For shallow quenches through I, the massless phase is metastable and the phase transition proceeds via the nucleation and growth of droplets of the massive phase. The merger of these droplets results in late-stage domain growth analogous to that for the unstable case. In all cases, the asymptotic growth process  exhibits dynamical scaling, and the growth law is $L(t)\sim t^{1/2}$. Given the dynamical universality of the processes involved, our results are of much under applicability than the simple NJL Hamiltonian considered here. We hope that our results will motivate fresh experimental interest in hot and dense quark matter, and that our predictions will be subjected to experimental verification.
\vspace{0.5cm}\\
{\bf Acknowledgments}\\
AS would like to thank CSIR (India) for financial support. HM would like to thank the School of Physical Sciences, Jawaharlal Nehru University, New Delhi for hospitality.

\end{document}